# Transport critical current, anisotropy, irreversibility fields and exponential *n* factors in Fe sheathed MgB$_2$ tapes

C. Beneduce**, H.L. Suo, P. Toulemonde, N. Musolino and R. Flükiger*

Département de Physique de la Matière Condensée, Université de Genève, 24 quai Ernest-Ansermet, CH-1211 Genève 4, Switzerland

***Abstract*** The influence of the initial MgB$_2$ grain size on critical current density, upper critical fields and irreversibility has been studied on Fe sheathed monofilamentary MgB$_2$ tapes prepared by the Powder-In-Tube technique. The effect of the reduction of MgB$_2$ grain size by ball milling was mainly to enhance both the critical current density, $j_c$, and the irreversibility field, $\mu_0 H_{irr}$, while the upper critical field, $\mu_0 H_{c2}$, remained unchanged. The anisotropy ratio of $\mu_0 H_{c2}$ between magnetic fields parallel and perpendicular to the tape surface was determined to 1.3, reflecting a deformation induced texture. A good agreement has been found between resistive and inductive $j_c$ values, measured at various temperatures. At 25K and 1 T, $j_c$ values close to $10^5$ A/cm$^2$ were measured. The exponential *n* factor of the resistive transition was found to be quite high at low fields, and decrease linearly from 60 at 4T to 10 at 8.5T.

∗ Corresponding author: Prof. René. Flükiger, Départment de Physique de la Matière Condensée, University of Geneva, 24 Quai Ernest Ansermet, 1211 Geneva 4, Switzerland. (Tel: +41-22-7026240, Fax: +41-22-7026869, E-mail: Rene.Flukiger@physics.unige.ch)

∗∗ Actual address: Bruker Biospin, Magnetics Division, Industriestrasse 26, Ch-8117 Fällanden, Switzerland

## Introduction

The discovery [1] of $MgB_2$ as a superconductor with a transition temperature of 39 K has attracted great scientific interest worldwide in the area of basic and applied research. Due to its intermediate transition temperature, weak-link free grain boundaries [2-3] and low material cost, this material has a potential for various electric power applications as well as for devices. Soon after its discovery, several research groups have tried to demonstrate the feasibility to fabricate $MgB_2$ wires or tapes with high critical current density using the Powder-In-Tube (PIT) method [4-11]. We have previously reported [4, 10] on the fabrication of highly dense monofilamentary $MgB_2$ tapes with large transport critical current density using Ni and Fe sheath materials. It was found that both the grain size of the starting powder and the appropriate deformation procedure as well as the heat treatment were determinant for getting higher $j_c$ values. In view of the application mentioned above, it is essential to study the transport behavior and the physical properties in $MgB_2$ tapes in order to understand the flux pinning mechanism and to further improve the tape properties. In this paper, we report on the irreversibility fields, electronic anisotropy, transport critical current densities and $n$ factors in those monofilamentary $MgB_2$ tapes. We found a significant effect of both the sheath material and initial $MgB_2$ powder grain size on transport $j_c$ values and upper critical fields.

## Experimental

The $MgB_2$ tapes were fabricated by sealing commercial $MgB_2$ powder into Fe or Ni tubes under an inert Ar atmosphere, followed by swaging, drawing and rolling. The powder used was a standard commercial $MgB_2$ powder from Alfa-Aesar with a purity of 98% which was ball milled for 2 hours down to smaller grain sizes (the powder size distribution is centred around 3 $\mu$m, showing agglomerates at 30 $\mu$m [10]). The Fe tubes had an outer diameter (OD) of 8 mm and an inner diameter (ID) of 5 mm, while the Ni tube had OD = 12.7 mm, ID = 7 mm. Both ends of the tubes were sealed with lead pieces. After drawing to a diameter of 2 mm, the wires were cold-rolled to a thickness of and 375 µm for $MgB_2$/Fe and 230 µm for $MgB_2$/Ni. These tapes were annealed for 0.5 h in a pure Ar atmosphere at 950 °C and 980 °C for the $MgB_2$/Fe and $MgB_2$/Ni tapes respectively. The details of the cross sections and microstructures of those tapes were reported elsewhere [4, 10].

The powder x-ray diffraction patterns of the crushed $MgB_2$ core revealed ~ 5 % of MgO as the only impurity phase after sintering. X-ray $\theta$-$2\theta$ diffraction patterns of a freshly etched surface from the outside part of the $MgB_2$ core, in contact with Fe, were also recorded at 1.5418 Å and the rocking curve around the (0 0 2) reflection was measured using a four circles diffractometer.

Inductive $j_c(H,T)$ values were calculated from the magnetic hysteresis loops of the $MgB_2$ cores with a standard Vibrating Sample Magnetometer (VSM), using the Bean critical state model for a semi - infinite slab. Magnetic fields up to 7.5T were applied parallel to the flat samples and M (H) loops were recorded every 5 K from T = 5 K to 35 K.

Critical current and resistance measurements as a function of temperature and applied magnetic field were performed on 45 mm long tape pieces in a He-flow cryostat using a four-probe technique. The temperature was measured on a current lead immediately next to the sample. The voltage contacts were 10 mm apart, and the voltage criterion used was 1 µV/cm. The magnetic field was perpendicular to the current direction and parallel to the tape surface, for all the critical current measurements. The electric resistance was measured using a d.c. current density of 10 A/cm$^2$ in fixed field applied perpendicular to the current direction and parallel to the tape surface ($H^{\parallel}$) or perpendicular to the tape surface ($H^{\perp}$), while sweeping the temperature. The upper critical fields were determined using the resisitive onset temperature, while the temperature at which the resistance vanishes was used for determining the irreversibility field.

## Results and discussions

**Effect of the sheath material**

Fig. 1 shows the field dependence of $j_c$ at 4.2 K in Fe/MgB$_2$ tapes (powder with and without ball milling) and Ni/MgB$_2$ tape (ball milled powder) after annealing. Lack of thermal stabilization causes sample quenching above a given current density. At 1.5 T, the Ni/MgB$_2$ tape had a transport $j_c$ of 2.3x10$^5$ a/cm$^2$. In a field of 6.5T its value was about 5 times lower than that of Fe/MgB$_2$ tapes (~10$^4$ A/cm$^2$) prepared using ball milled powder. The reason of this difference can be understood by the variation of the resistive superconducting transition under magnetic fields parallel to the tape surface plotted in Fig. 2. In zero field the Fe/MgB$_2$ tape has a T$_c$ of 37.3K with $\Delta$T$_c$= 0.4K, while Ni/MgB$_2$ has a T$_c$ of 36.5K with $\Delta$T$_c$=0.8K, the T$_c$ difference becoming bigger at higher fields. The $H_{c2}(T)$ curves show a positive curvature near T$_c$ but are linear at intermediate temperatures, the slopes being d$H_{c2}$/d$T$ = -0.57 T/K and - 0.33 T/K for Fe/MgB$_2$ and Ni/MgB$_2$ tapes, respectively. The main effect of the Ni sheath is to lower the upper critical field as shown in Fig. 2. As mentioned earlier [4], the difference in the transport $j_c$ values of the annealed MgB$_2$/Ni and MgB$_2$/Fe tapes is mostly due to chemical reasons. In particular, the reaction between Ni and MgB$_2$ had a negative effect on the core with a decrease of the transport properties.

**Effect of the powder grain size**

To study the effect of the initial powder grain size on the transport critical current density, we prepared Fe/MgB$_2$ tapes directly from the as-purchased powder (Alfa-Aesar), and with ball milled powder. The deformation and heat treatment was the same as the one described above. The crashing process led to a decrease of the average grain size. As reported previously [10], the as–purchased powder contains a large number of agglomerated grains, with a wide size distribution centred at around 60 µm. After 2 hours of ball milling, 35% of the grains have diameters around 3 µm, while 60% are centred around 30 µm.

The transport critical current density for annealed Fe/MgB$_2$ tape prepared by commercial powder is also shown in Figure. 1, a transport critical current density of $4 \times 10^4$ A/cm$^2$ at 4.2 K in a field of 3.75 T was obtained. In a field of 6.5 T, this tape had a $j_c$ value of $2.4 \times 10^3$ A/cm$^2$, i.e. 4 times less compared to $10^4$ A/cm$^2$ for the sample prepared starting from ball-milled powder.

Figure. 3 (a) and (b) compare the SEM micrographs of both polished MgB$_2$ cores in Fe/MgB$_2$ tapes prepared by commercial powder and ball milled powder, respectively. A rough and porous microstructure of MgB$_2$ core is observed in tape produced by the as-received commercial powder. The high porosity and poor microstructure in the MgB$_2$ core of this wire is likely due to the large grain size of the powder and to its surface state, leading to a degradation of the transport $j_c$ value in this tape, as mentioned above. This is in contrast to the relatively flat and dense aspect of MgB$_2$ cores of tape made by ball milled powder. This important improvement of the MgB$_2$ core quality is believed to be responsible for the higher transport $j_c$ in this tape.

The temperature dependence of the upper critical field $\mu_0 H_{c2}$ and of the irreversibility field $\mu_0 H_{irr}$ for the Fe sheathed tapes with commercial and ball-milled powder are shown in Fig. 4. The magnetic field was applied parallel to the tape surface. Compared to the tape with commercial powder, the ball-milled tape exhibited much higher irreversibility fields, especially at low temperatures, whereas $\mu_0 H_{c2}$ data of the two samples are comparable. The shift of the irreversibility line and the enhancement of the critical current density of the ball milled tape are due to improved flux pinning and may be caused by the higher number of grain boundaries. Similar effect of the grain size on the irreversibility line was found in bulk untextured samples [12] and MgB$_2$ films [13].

**Electronic anisotropy**

Since MgB$_2$ consists of alternating B and Mg hexagonal sheets, electronic anisotropy should be expected. Reported values of $\gamma = \dfrac{\mu_0 H_{c2}^{//}(0K)}{\mu_0 H_{c2}^{\perp}(0K)}$ span between 1 [14] and 13 [15-16]. These values were obtained on MgB$_2$ in different form and sample quality. Experiments performed on randomly oriented powder samples [17], aligned MgB$_2$ crystallites [18] or single crystals [19-22] have given values from 6, 3±0.2 and 2.6-3 respectively. A somewhat smaller anisotropy of 1.8-2.5 has also been observed in $c$-axis oriented thin films [23-25]. Since MgB$_2$ is a highly promising candidate for technological applications, it is important to resolve the issue of magnitude of anisotropy in this system. Because the reports on different forms of MgB$_2$ are drastically different, it is useful to examine the anisotropic nature of our annealed Fe/MgB$_2$ tapes.

In order to explore the anisotropic behavior of the Fe/MgB$_2$ tapes, we measured the superconducting transition under magnetic field up to 14 T. Fig. 5 shows the resistive transitions with the field applied parallel (upper panel) and perpendicular (lower panel) to the tape surface. Increasing the magnetic field both onset and offset temperatures shift to lower temperature, and the superconducting transition becomes slightly broader. The field-induced decreases of T$_c$ were larger for the perpendicular configuration than for the parallel configuration, and the transition width for parallel fields was narrower than that for perpendicular fields (transition widths at 8 T are 3.2 K and 8.2 K for the parallel and perpendicular configurations, respectively). In Fig. 6, we report $\mu_0 H_{c2}$ as a function of temperature for both orientations: the upper critical field is considerably higher when the field is perpendicular to the c-axis (i.e. parallel to the tape surface). As in other reports [26], the $H_{c2}(T)$ curve shows a positive curvature near T$_c$. We evaluated d$H_{c2}$/d$T$ of the $\mu_0 H_{c2}$ -T curves for fields between 2 and 14T, and the slopes are - 0.45 T/K and - 0.57 T/K for the perpendicular and the parallel configuration, respectively. By using the dirty limit extrapolation we found $\mu_0 H_{c2}^{//} = 15.1$T and $\mu_0 H_{c2}^{\perp} = 11.9$T. From these data it is possible to calculate the temperature independent anisotropy factor in our tape to γ = 1.3. The same value for the anisotropic factor was found in Ta/MgB$_2$ wires [27]. This value is smaller than that for the c-axis oriented thin films (1.8-2.2) [23-24] and than the anisotropy factor of MgB$_2$ single crystals (2.6) [22]. This result suggests a possible alignment of the grains along a crystallographic axis, i.e. a preferential orientation of the core grains, especially at the interface with the iron sheath. Structural Rietveld refinements were done on XRD patterns of the MgB$_2$ core at the Fe sheath interface. Figure 7 shows the difference of the calculations obtained considering only an isotropic repartition of the crystallites and taking into account the MgO and FeB$_2$ (as observed in Fe doped MgB$_2$ samples [28]) impurity phases and the refinement of the Mg site occupancy. In the first case, the under-estimation of the (0 0 l) peaks (see (0 0 2) peak in Fig. 7) shows that a preferential orientation along the c-axis exists at the interface with the Fe sheath. In the second case, the (0 0 l) reflections are well fitted using a platelet model for the grains. Moreover, the refinement shows the presence of ~ 10 % of vacancies on the Mg site, with no significant variation of the lattice parameters (a = 3.0835(5) Å, c = 3.524(3) Å) compared to the stoichiometric MgB$_2$ (a = 3.0834(3) Å, c = 3.5213(6) Å [29]). This non-stoichiometry on the Mg site, as the possible lattice strain [30], could explain the lower $T_c$ = 37.3 K value measured (see the resistance curve at 0 T in Fig. 5), as previously observed by Serquis et al. [30]. The rocking curve around the (0 0 2) reflection gives a FWHM of ~ 18°, which confirms the slightly preferential orientation along the c-axis of the MgB$_2$ grains near the Fe sheath, but this high value also shows that the texture is not very pronounced. Further optimization of the deformation process is expected to give higher anisotropy factors.

**Transport and inductive $j_c$ measurements**

Most of the transport $j_c$ results reported so far on $MgB_2$ wires and tapes are limited at 4.2K. To take advantage of the relatively high critical temperature it is important to have high $j_c$ values at temperatures above 20 K. Figure 8 shows both inductive (solid lines) and transport (symbols) $j_c$ values of $MgB_2$/Fe tape, measured every 5 K between 10 and 35 K and plotted against the magnetic field. Below $j_c \approx 10^3$ A/cm$^2$, the inductive data start to get noisy, while above $j_c \approx 10^4$ A/cm$^2$ the transport measurements start to be affected by sample quenches due to the insufficient thermal stability, the electrical resistivity of Fe being too high. The temperature of the sample was monitored during the measurement, showing a temperature rise of approximately 0.5K at high currents. The transport $j_c$ values are consistent with those obtained from magnetisation measurements. The field dependence of $j_c$ is essentially the same for transport and magnetisation, well above the self field regime $j_c$ decays approximately exponentially, with a characteristic decay field which is very similar in the inductive and transport data. The transport $j_c$ values are higher than those of the magnetic ones at fixed field and temperature and can be understood in terms of the different criteria fixed for the two independent experiments. In the transport measurements this criterion was $E_{c,\text{transport}} = 10^{-6}$ V/cm, whereas in the magnetisation experiments the induced electric field can be estimated to be ~ 100 times lower : $E_{c,\text{inductive}} \approx 10^{-8}$ V/cm. At T = 25 and 30 K, transport $j_c$ well above $10^4$ A/cm$^2$ were obtained at fields of 2.25 and 1.0 T, respectively.

**The exponential $n$ factor**

The upper critical field and the irreversibility field for $MgB_2$ are well above those of NbTi. As several groups are making rapid progress in producing $MgB_2$ based wires or tapes, the prospect of low-cost superconducting solenoids based on this compound becomes ever more realistic. To use such magnets in persistent mode the *n* factor of the wire should have a value greater than 30 at the operation field [31-32].

The logarithmic *EJ* curves measured at 4.2 K can be reasonably well approximated by a local power-law around our electric field criterion $E_c = 10^{-5}$ V/cm: $\left(\frac{E}{E_c}\right) \approx \left(\frac{j}{j_c}\right)^n$. By fitting this relation to our data in the electric field range $5 \times 10^{-6}$ V/cm $< E < 5 \times 10^{-5}$ V/cm, we obtain the factors shown in Figure 9. The curve has an exponential magnetic field dependence. The factor is determined to be ~ 30 at 6 T, and decreases exponentially to 10 at 8.5 T.

The high value of the factors at relatively high magnetic field has an important practical consequence, since it will be possible to use such magnets in persistent mode for fields smaller than 6 T.

## Conclusion

We report on transport critical current, anisotropy, irreversibility fields of $MgB_2$/Fe and $MgB_2$/Ni tapes fabricated by a Powder-In-Tube (PIT) technique. It is found that the upper critical field of annealed Fe/$MgB_2$ tapes is slightly improved with respect to the bulk values, while a decrease is observed in $MgB_2$/Ni tapes due to the reaction between Ni and the $MgB_2$ core. The study on the effect of the initial $MgB_2$ powder grain size shows that both the critical current density and the irreversibility field are enhanced in $MgB_2$/Fe tape produced by ball milled powder, while the upper critical field, $\mu_0 H_{c2}$, remained unchanged. The transport $j_c$ values are consistent with those obtained from magnetization measurements. Transport $j_c$ values above $10^4$ A/cm$^2$ are obtained in $MgB_2$/Fe tapes at 4.2 K/6.5 T, 25 K/2.25 T and 30 K/1 T. For higher currents, however, these monofilamentary tapes quenched due to insufficient thermal stability. Thus, a self-field transport $j_c$ value in this tape at 4.2 K can only be extrapolated, yielding a value close to $10^6$ A/cm$^2$ (confirmed by inductive measurements).

Electronic anisotropy measurements indicate an upper critical field anisotropy ratio of 1.3. Our tapes exhibit a very high *n* factor, i.e. 43 at 5 T, which opens possibilities for a persistent mode operation. Further enhancement of $j_c$ in tapes is expected for high degrees of texture. Particular effort will be done in view of the decrease of to filament thickness and improvement of the thermal stability of the sample.

## Acknowledgment

We thank Robert Janiec for his help with the transport measurements, Patrick Cerutti and Aldo Naula for their technical support in the tape annealing and Dr. XiaoDong Su for his help in the tape preparation. This work was supported by the Fond National Suisse de la Recherche Scientifique.

## References

[1] J. Nagamatsu, N. Nakagawa, T. Muranka, Y. Zenitanim and J. Akimitsu, Nature, **410** 63 (2001).

[2] D.C. Larbalestier, L. D. Cooley, M. O. Rikel, A. A. Polyanskii, J. Jiang, S. Patnaik, X. Y. Cai, D.M. Feldmann, A. Gurevich, A. A. Squitieri, M. T. Naus, C. B. Eom, E. E. Hellstrom, R. J. Cava, K. A. Regan, N. Rogado, M. A. Hayward, T. He, J. S. Slusky, P. Khalifah, K. Inumaru and M. Haas, Nature, **410,** 186 (2001).

[3] M. Dhallé, P. Toulemonde, C. Beneduce, N. Musolino, M. Decroux, and R. Flükiger, Physica C. **363**, 155 (2001).

[4] H.L. Suo, C. Beneduce, M. Dhallé, N. Musolino, J-Y. Genoud and R. Flükiger. Appl. Phys. Lett. **79**, 3116 (2001).

[5] W Goldacker, S I Schlachter, S Zimmer and H Reiner, Supercond. Sci. Technol. **14** 787-793 (2001)

[6] B. A. Glowacki, M. Majoros, M. Vickers, J. E. Evetts, Y. Shi, and I. Mcdougall, Supercond. Sci. Technol. **14,** 193 (2001).

[7] S. Jin, H. Mavoori, C. Bower and R. B. van Dover, Nature **411**, 563 (2001).

[8] S. Soltanian, X.L. Wang, I. Kusevic, E. Babic, A.H. Li, H.K. Liu, E.W. Collings and S.X. Dou, Physica C **361**, 84-90 (2001).

[9] G. Grasso, A. Malagoli, C. Ferdeghini, S. Roncallo, V. Braccini, M. R. Cimberle and A. S. Siri, Appl. Phys. Lett. **79**, 230 (2001).

[10] H.L. Suo, C. Beneduce, M. Dhallé, N. Musolino X.D. Su, E. Walker and R. Flükiger. CEC/ICMC conference, Madison, USA July (2001), accepted for publication in "Advances in Cryogenic Engineering".

[11] H. Kumakura, A. Matsumoto, H. Fujii and K. Togano, Appl. Phys. Lett. **79**, 2435 (2001).

[12] A. Gümbel, J. Eckert, G. Fuchs, K. Nenkov, K.-H. Müller, L. Schultz, cond-mat/0111585

[13] C.B. Eom, M.K. Lee, J.H. Choi, L. Belenky, X. Song, L.D. Cooley, M.T. Naus, S. Patnaik, J. Jiang, M. Rikel, A. Polyanskii, A. Gurevich, X.Y. Cai, S.D. Bu, S.E. Babcock, E.E. Hellstrom, D.C. Labalestier, N. Rogado, K.A. Regan, M.A. Hayward, T. He, J.S. Slusky, K. Inumaru, M.K. Haas and R.J. Cava, Nature **411**, 558 (2001).

[14] X. H. Chen, Y.Y. Xue, R.L. Meng, C. W. Chu, Phys. Rev. B **64**, 172501 (2001).

[15] S. R. Shinde, S. B. Ogale, A. Biswas. R. L. Greene, T. Venkatesan, cond-mat 0110541.

[16] F. Simon, A. Jánossy, T. Fehér, and F. Murányi, Phys. Rev. Lett., vol **87**, 047002 (2001).

[17] S. L. Bud'ko, V. G. Kogan, and P. C. Canfield, Phys. Rev. B **64**, 180506 (2001).

[18] O. F. de Lima, R. A. Ribeiro, M. A. Avila, C. A. Cardoso, and A. A. Coelho Phys. Rev. Lett., vol **87**, 5974, (2001).

[19] S. Lee, H. Mori, T. Masui, Y. Eltsev, A. Yamamoto and S. Tajima, J. Phys. Soc. Jpn. **70**, 2255 (2001).

[20] A. K. Pradhan, Z. X. Shi, M. Tokunaga, and T. Tamegai, Y. Takano and K. Togano, H. Kito and H. Ihara, Phys. Rev. B **64**, 212509 (2001)

[21] Kijoon H. P. Kim, Jae-Hyuk Choi, C. U. Jung, P. Chowdhury, Hyun-Sook Lee, Min-Seok, Park, Heon-Jung Kim, J. Y. Kim, Zhonglian Du, Eun-Mi Choi, Mun-Seog Kim, W. N. Kang, Sung-Ik Lee, Gun Yong Sung, Jeong Yong Lee, Phys. Rev. B **65**, 100510 (2001).

[22] M. Xu, H. Kitazawa, Y. Takano, J. Ye, K. Nishida, H. Abe, A. Matsushita, N. Tsujii, G. Kido, Appl. Phys. Lett. **79**, 2779 (2001)

[23] C. Ferdeghini, V. Ferrando, G. Grassano, W. Ramandan, E. Bellingeri, V. Braccini, D. Marré, P. Manfrinetti, A. Palenzona, F. borgatti, R. Felici, T-L Lee, Supercond. Sci. Technol. **14**, 952 (2001)

[24] S. Patnaik, L. D. Cooley, A. Gurevich, A. A. Polyanskii, J. Jiang, X.Y. Cai, A.A. Squitieri, M.T. Naus, M.K. Lee, J. H. Choi, L. Belenky, S. D. Bu, J. Letteri, X. Song, D. G. Schlom, S. E. Babcock, C. B. Eom, E. E. Hellstrom, D. C. Larbalestrier, Supercond. Sci. Technol. **14**, 315 (2001)

[25] R. Vaglio, M.G. Maglione, R. Di Capua cond-mat/0203322

[26] P.C. Canfield, D.K. Finnemore, S.L. Bud'ko, J.E. Ostenson, G. La-Pertot, C.E. Cunningham and C. Petrovic, Phys. Rev. Lett., vol **87**, 2423, (2001)

[27] A. K. Pradhan, Y. Feng, Y. Zhao, N. Koshizuka, L. Zhou, P.X. Zhang, X. H. Liu, P. Ji, S.J. Du, C. F. Liu, Appl. Phys. Lett **79**, 1649 (2001).

[28] M. Kühberger and G. Gritzner, to be published in Physica C.

[29] M.E. Jones, R.E. Marsh, Journal of the American Chemical Society **76**, 1434-1436 (1954), Acta Crystallographica **1**, 1948-23 (1967).

[30] A. Serquis, Y. T. Zhu, E. J. Peterson, J. Y. Coulter, D. E. Peterson, and F. M. Mueller, Appl. Phys. Lett **79**, 4399 (2001).

[31] B. Seeber in "*Handbook of Applied Superconductivity*", B. Seeber Ed., Inst. Of Phys. Publishing, Bristol, 307 (1998)

[32] W.H. Tschopp, D.D. Laukien in "*Handbook of Applied Superconductivity*", B. Seeber Ed., Inst. Of Phys. Publishing, Bristol, 1191 (1998)

**Figure captions**

**Figure 1.** Transport critical current densities at T = 4.2 K as function of applied field in annealed Fe (with and without ball milling) and Ni (with ball milling) sheathed $MgB_2$ tapes.

**Figure 2.** The upper critical fields versus temperature for both annealed Fe and Ni sheathed $MgB_2$ tapes.

**Figure 3.** SEM microstructures of the Fe/$MgB_2$ tapes with and without ball milling : (a) $MgB_2$ core in the annealed Fe/$MgB_2$ tapes prepared by commercial powder ; (b) $MgB_2$ core in the annealed Fe/$MgB_2$ tape prepared by ball milled powder.

**Figure 4.** The temperature dependence of the upper critical field $\mu_0H_{c2}$ and of the irreversibility field $\mu_0H_{irr}$ for the Fe-sheathed tapes prepared by commercial and ball milled powder.

**Figure 5.** The resistive transitions with the field applied parallel (upper panel) and perpendicular (lower panel) to the sample surface.

**Figure 6.** The upper critical field, $\mu_0H_{c2}$ as a function of temperature for field perpendicular and parallel to the sample surface in annealed $MgB_2$/Fe tape.

**Figure 7.** Rietveld refinement profile of x-ray diffraction pattern ($\lambda$ = 1.542 Å) of the outside part of the $MgB_2$ core, at the interface with the Fe sheath. A difference curve is plotted at the bottom (observed minus calculated). Tick marks correspond to Bragg peaks of $MgB_2$, MgO and $Fe_2B$. Inset : Rietveld refinement without taking into account any impurity phases, any preferential orientation of the $MgB_2$ grains and a full Mg site.

**Figure 8.** The field dependence of $j_c$ for annealed $MgB_2$/Fe tape, measured every 5 K between 10 and 35 K either inductively (solid lines) or by transport experiments (symbols).

**Figure 9.** The field dependence of the n-factors in $MgB_2$/Fe tape.

Figure 1

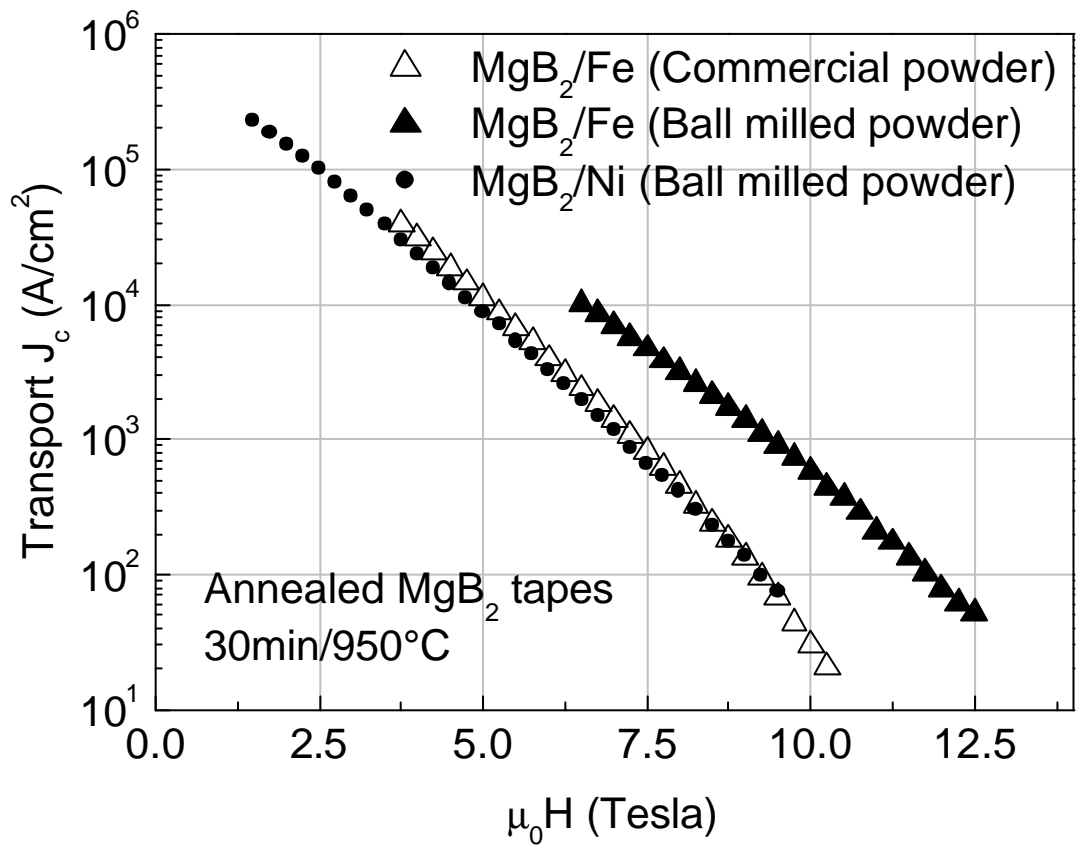

Figure 2

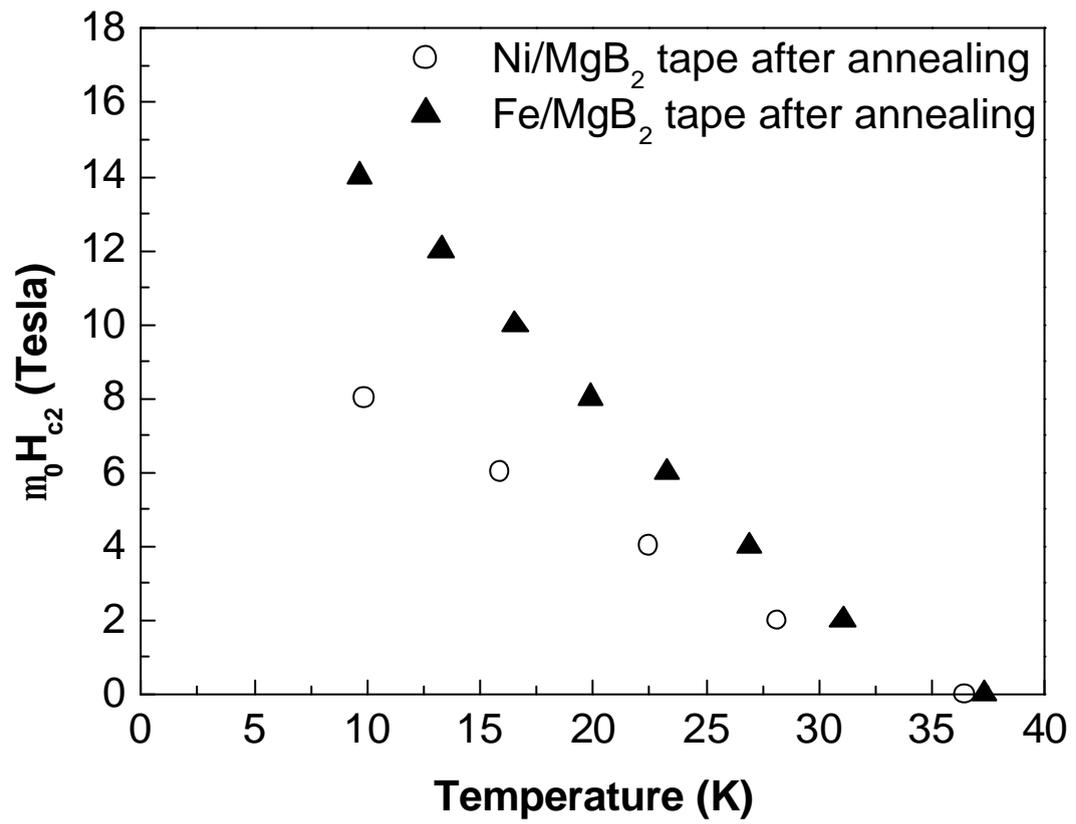

Figure 3

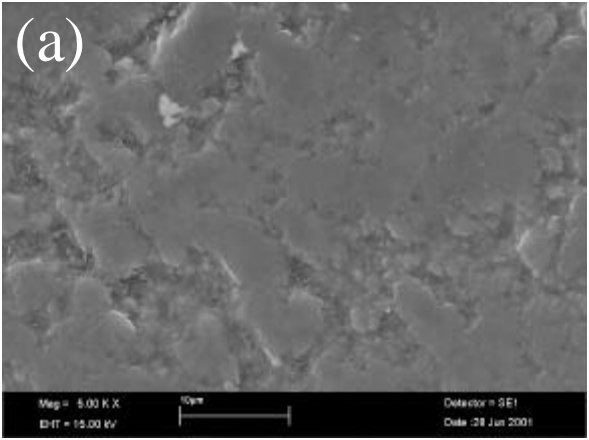
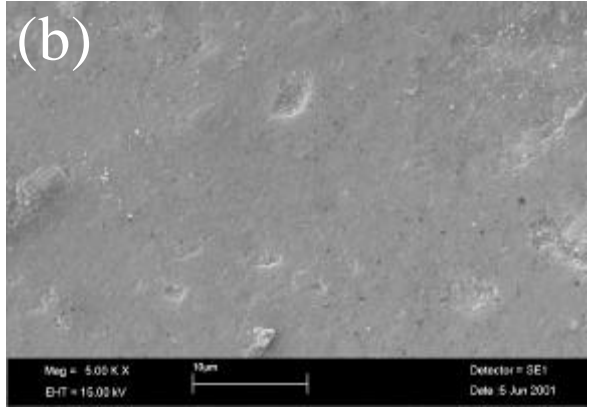

Figure 4

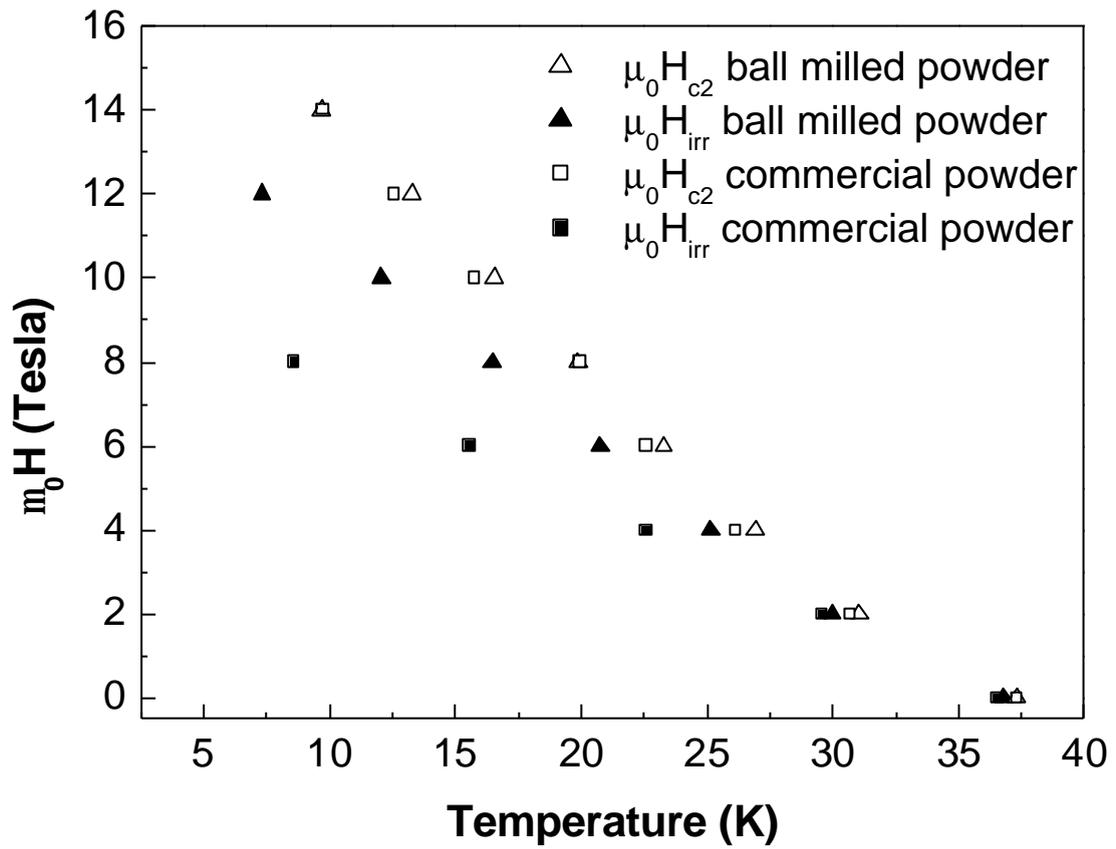

Figure 5

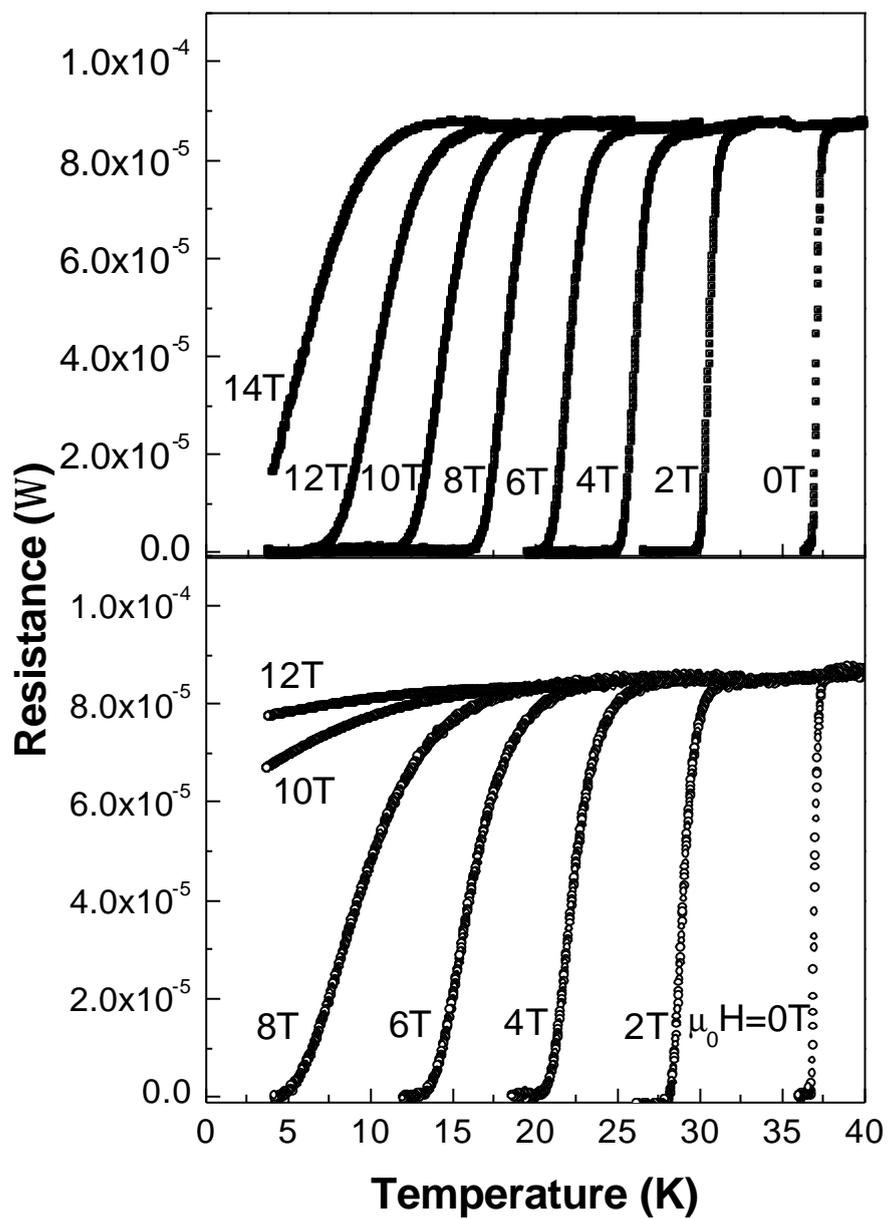

Figure 6

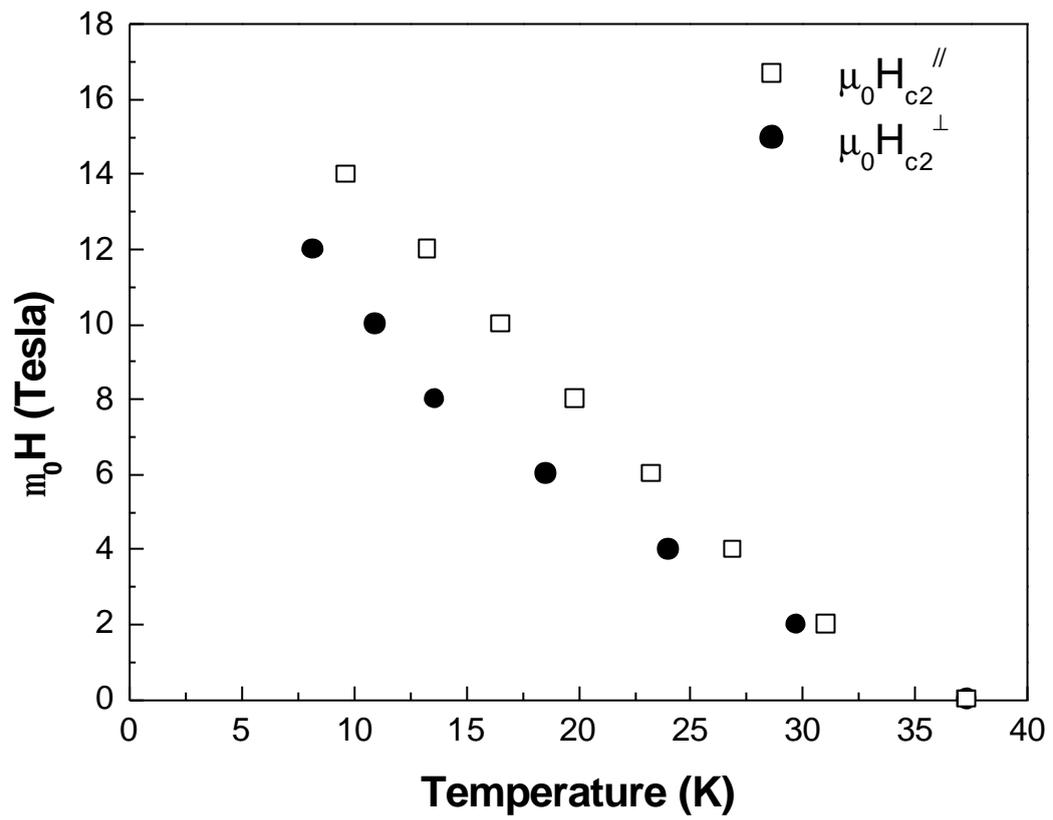

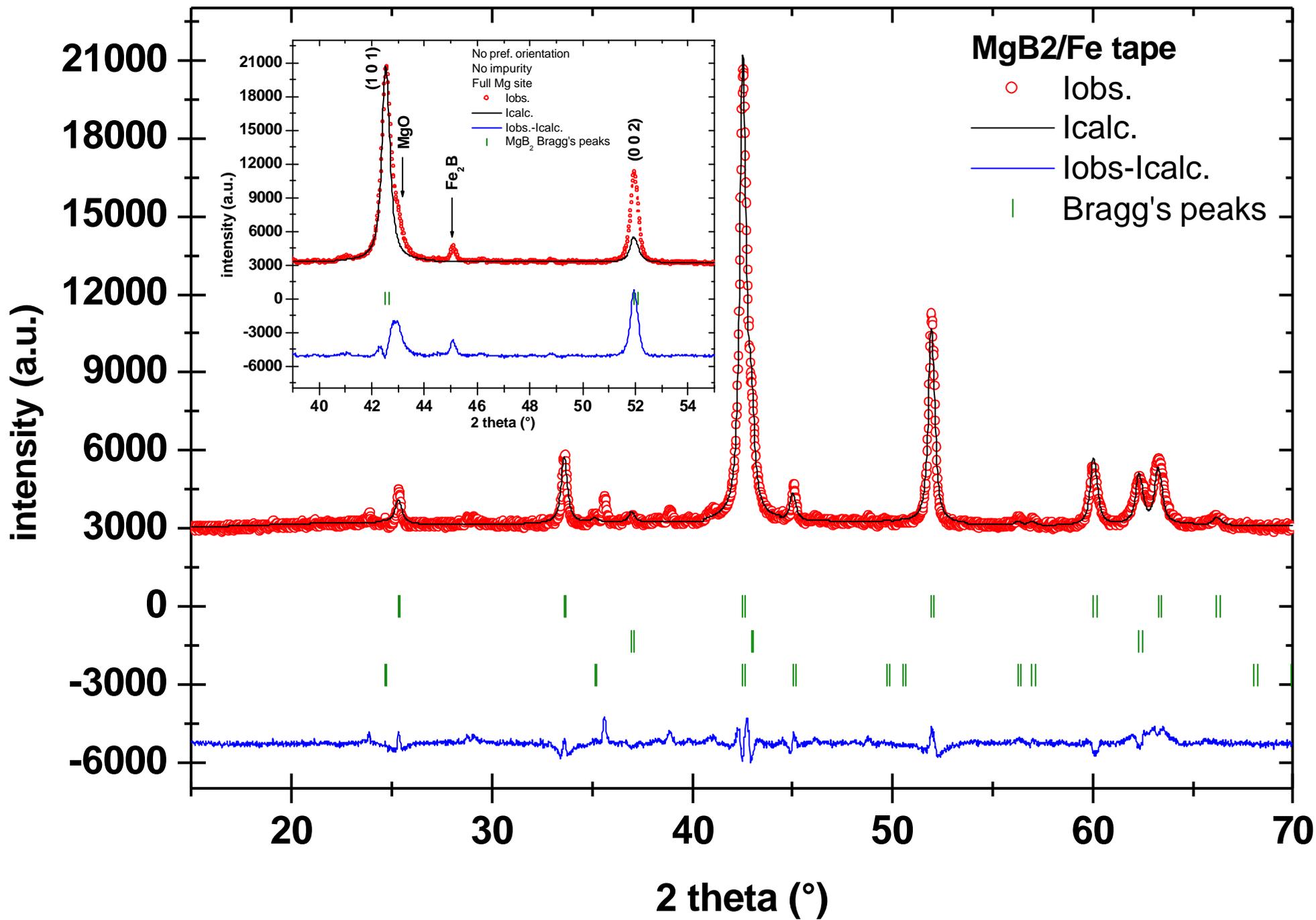

Fig. 7.

Figure 8

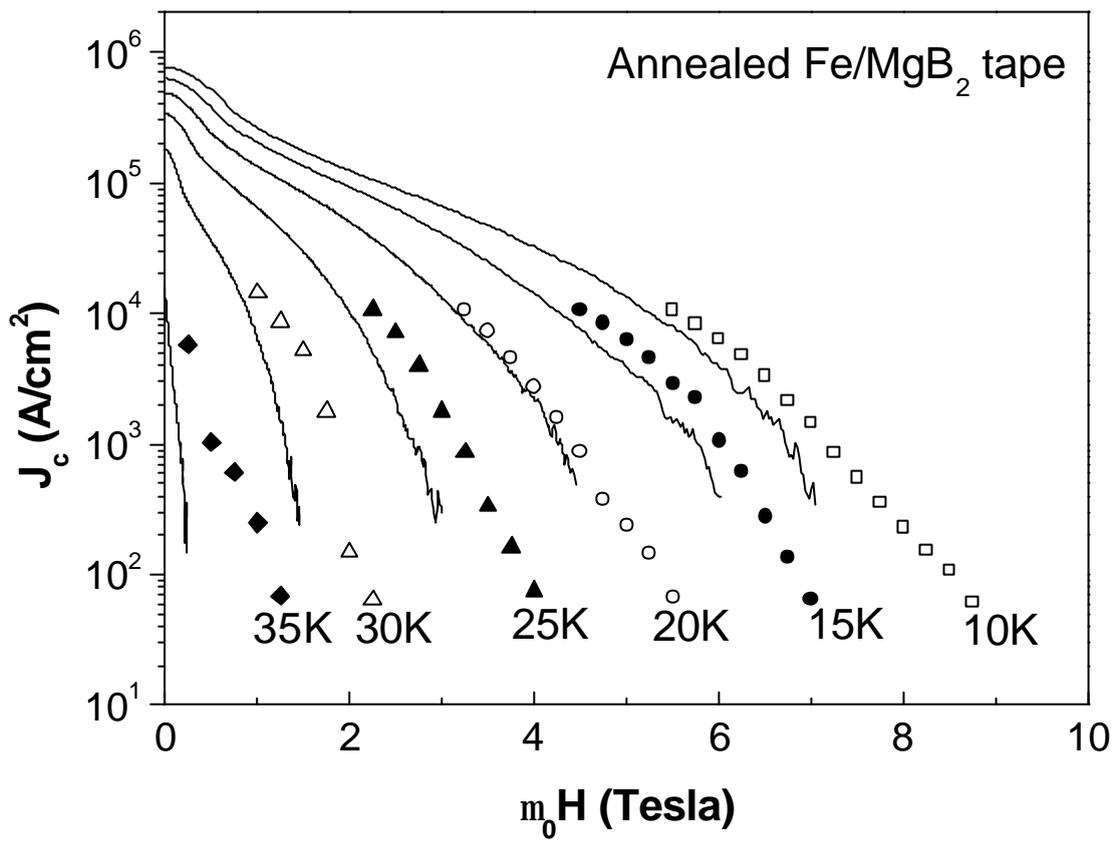

Figure 9

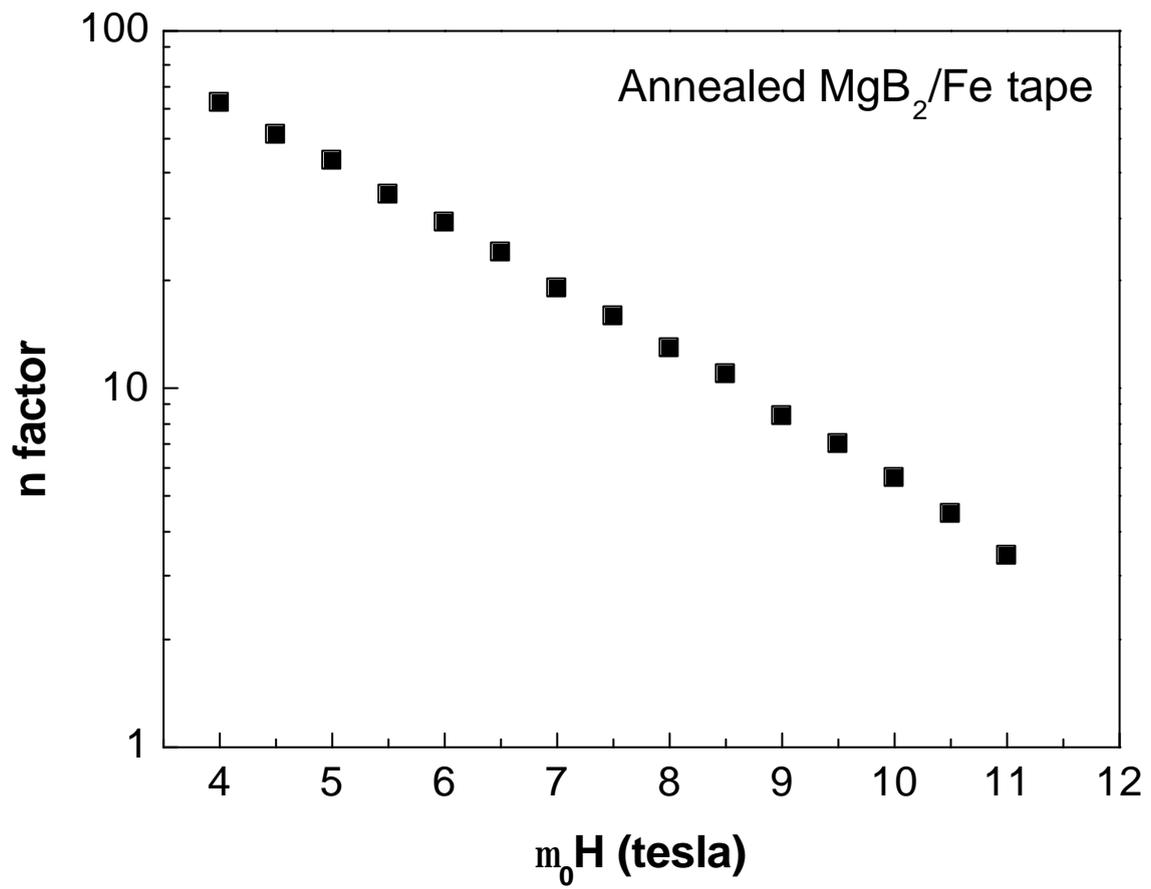